\documentclass[review]{elsarticle}

\usepackage{epsfig}
\usepackage{amssymb}

\usepackage{color}
\definecolor{Red}{rgb}{0.9,0.0,0.1}

\usepackage{amsmath}
\usepackage{amsfonts}
\usepackage{multirow}
\usepackage{graphicx,psfrag,epsf}
\usepackage{enumerate}
\usepackage{natbib}
\usepackage{url} 

\usepackage{lineno,hyperref}
\modulolinenumbers[5]

\journal{Mathematics and Computers in Simulation}

\begin{document}
\begin{frontmatter}

\title{Linear-Phase-Type probability modelling of functional PCA with applications to resistive memories}


\author[mymainaddress]{Juan E. Ruiz-Castro\corref{mycorrespondingauthor}}
\cortext[mycorrespondingauthor]{Corresponding author}
\ead{jeloy@ugr.es}

\author[mymainaddress]{Christian Acal}
\address[mymainaddress]{Department of Statistics and O.R. and IEMath-GR, University of Granada, Spain}

\author[mymainaddress]{Ana M. Aguilera}

\author[mysecondaryaddress]{M. Carmen Aguilera-Morillo}
\address[mysecondaryaddress]{Department of Applied Statistics and Operational Research, and Quality,
Universitat Polit\`{e}cnica de Val\`{e}ncia and 
UC3M-BS Santander Big Data Institute, Spain}

\author[mytertiaryaddress]{Juan B. Rold\'an}
\address[mytertiaryaddress]{Department of Electronics and Computer Technology. University of Granada, Spain}

\begin{abstract}
Functional principal component analysis (FPCA) based on Karhunen-Lo\`eve (K-L) expansion  allows to describe the stochastic evolution of the main characteristics associated to multiple systems and devices. Identifying the probability distribution of the principal component scores is fundamental to characterize the  whole process. The aim of this work is to consider a family of statistical distributions that could be accurately adjusted to a previous transformation. Then, a new class of distributions, the linear-phase-type, is introduced to model the {\color{Red}{principal components.}} This class is studied in detail in order to prove, through the K-L expansion, that certain linear transformations of the process at each time point are phase-type distributed. This way, the one-dimensional distributions of the process are in the same linear-phase-type class. Finally, an application to model the reset process associated with resistive memories is developed and explained.
\end{abstract}


\end{frontmatter}


\section{Introduction}

Among the electron devices with greater potential in the current microelectronic industry landscape are Resistive Random Access Memories (RRAMs). The number of indexed publications in this field has skyrocketed and therefore, the attention of the academic community as well as the electronics companies' development teams is fixed on them. The applications of these new devices range from non-volatile memory circuits, security modules for cryptography and neuromorphic computation \cite{Pan14}.

 The stochastic nature of the physical mechanisms behind RRAM resistive switching (RS) operation makes the statistical modelling of the inherent device stochasticity essential. The key issue here {\color{Red}{rests}} upon the need to correctly explain variability in the current/voltage curves associated with long series of successive RS cycles \cite{Aguilera-Morillo19, Perez19, Roldan19, Acal19}, i.e., cycles of continuous reset and set processes. If the device charge conduction is filamentary, the most common case, RS cycles get translated into rupture and rejuvenation of conductive filaments that dramatically changes the device resistance \cite{G.Cordero17}. The modelling of the current versus voltage curves in these devices is of most importance for circuit design. Therefore, in this context, and taking into consideration that the experimental data we have are curves, an approach based on functional data analysis (FDA) can be applied in order to accurately model resistive memory characteristics.

A deep description of the main FDA methods with applications in different fields was developed in \cite{Ramsay05}. Functional principal component analysis (FPCA) based on Karhunen-Loève (K-L) expansion provides an orthogonal representation of an stochastic process in terms of uncorrelated  random variables, called principal components (p.c.'s). The K-L expansion can be truncated so that the process is approximated  in terms of the most explicative p.c.'s \cite{Aguilera13a}. A three step algorithm for estimating FPCA from the reset curves (current versus voltage curves) of a sample of RRAM cycles was proposed in \cite{Aguilera-Morillo19}. This new type of modelling can be very attractive from the circuit simulation viewpoint because it allows to describe the main characteristics of these devices, such as variability. Making use of this technique, the implementation of variability in  compact models for RRAMs can be greatly simplified.

Nevertheless, identifying  the probability distribution of the principal components is fundamental to characterize the whole process through the K-L expansion. In previous studies, several authors have considered different transformations and used them as a starting point, they have fitted different distributions successfully. However, to find the appropriate transformation and its probability distribution is not an easy task. The aim of this work is to consider a family of statistical distributions that could be accurately adjusted for any transformation. In this respect, a new methodology is developed by considering phase-type distributions {\color{Red}{(PH)}} that were applied in \cite{Acal19, Perez19} for modelling the reliability functions associated to RRAM reset points, among others parameters. This class of distributions have been also considered in other multiple science fields such as queueing theory and reliability (\cite{Ruiz-Castro18}, \cite{Ruiz-Castro20a}, \cite{Ruiz-Castro20b}). The properties of this distribution class are very interesting and allow to achieve results in a well structured form. The developments and results can be expressed in an matrix-algorithmic and computational way. One of the main advantages of this class is that any non-negative distribution can be approximated as needed through a {\color{Red}{PH}} distribution \cite{Neuts81}. In order to fit  this distribution, the p.c.'s scores should be transformed previously to positive values. In fact, in this research, it is proved that for several transformations, the fit obtained is more accurate by considering {\color{Red}{PH}} distributions than any other distribution. A new class of distributions are introduced, the {\color{Red}{linear}}-phase-type distributions (LPH) defined as variables for which there is a linear transformation  that is {\color{Red}{PH}} distributed. This class is studied in detail in order to prove, through K-L expansion, that certain linear transformations of the process at each time point is {\color{Red}{PH}} distributed too.

In addition to this introduction, the paper has three other sections. The new LPH distributions and their main properties are studied in detail in Section 2. Then, the one-dimensional LPH distributions of the process are obtained from the LPH distributions of the p.c.'s through the K-L expansion. Finally, the proposed methodology is applied on different samples of current/voltage curves associated to RRAM devices in Section 4.

\section{{\color{Red}{LPH}} modelling}

In reliability, computer and electronic engineering, physics, queues theory and other fields, multiple probability distributions are frequently used, including the exponential, Erlang and Weibull distributions. Most of them involve calculations that may become unmanageable, due to the analytic expressions required. {\color{Red}{PH}} play an important role in this respect. This type of distributions enables us to express the main results in an algorithmic and computational way. This class of distribution was described in detail in \cite{Neuts81}.

\subsection{PH distributions}

\noindent \textbf{Definition 1} A nonnegative random variable $X$ is a {\color{Red}{PH}} distribution if its reliability function is given by
\[
R\left( x \right) = P\left\{ {X > x} \right\} = {\boldsymbol{\alpha }} e^
{{\bf{T}}x}  {\bf{e}} \ \ \ ; \ \ \ x \ge 0,
\]
 where $\boldsymbol{\alpha }$ is a substochastic vector of order \textit{m}, \textbf{T} a subgenerator of order \textit{m} (matrix $m\times m$ where all diagonal elements are negative, all off-diagonal elements are non-negative, invertible and all row sums are non-positive) and{\color{Red}{, throughout the paper,}} \textbf{e} is a column vector of ones with appropriate order.\\

 A {\color{Red}{PH}} distribution can be defined as the time up to the absorption in an absorbent Markov chain with initial distribution and generator for the transient states $\boldsymbol{\alpha}$ and \textbf{T}, respectively. In this case, $(\boldsymbol{\alpha}, \textbf{T})$  is called the representation of the {\color{Red}{PH}} distribution.\\

 Multiple good properties of these distributions are described  in \cite{Neuts81}. One of the main properties is that of {\color{Red}{PH}} distributions can approximate arbitrarily closely any probability distribution defined on the nonnegative real line.

\subsection{{\color{Red}{LPH}} distributions}

 A new probability distribution class is defined in this subsection. This class is called the {\color{Red}{linear}}-phase-type distribution class (LPH). A LPH distribution is defined as follows.\\

\noindent \textbf{Definition 2} A random variable $X$ follows a {\color{Red}{LPH}} distribution if $Y=a+bX$ is {\color{Red}{PH}} distributed for $a$ and $b$ ($b\neq 0$) in $\mathbb{R}$.\\

 If the representation of $Y$ is  $(\boldsymbol{\alpha}, \textbf{T})$ then the reliability function of $X$ (LPH) is
\[
{R_X}\left( x \right) = P\left( {X > x} \right) = \left\{
{\begin{array}{*{20}{l}}
	{{\boldsymbol{\beta }}{e^{{\bf{S}}x}}{\bf{e}}}&;&{{\rm{for}}\ x > \frac{{ - a}}{b};b >
		0}\\
	{1 - {\boldsymbol{\beta }}{e^{{\bf{S}}x}}{\bf{e}}}&;&{{\rm{for}}\ x < \frac{{ - a}}{b};b
		< 0}
	\end{array}} \right.,
\]
where $\boldsymbol{\beta}=\boldsymbol{\alpha}e^{\textbf{T}a}$, $\textbf{S}=b\textbf{T}$ and \textbf{e} is a column vector with appropriate order. In this case, will we denote the 4-tuple $(a,b,\boldsymbol{\beta},\textbf{S})$ as the representation of the corresponding LPH.

The density function of this class of distributions is given by
\[
{f_X}\left( x \right) = \left\{ {\begin{array}{*{20}{l}}
	{ - {\boldsymbol{\beta }}{e^{{\bf{S}}x}}{{\bf{S}}^0}}&;&{{\rm{for}}\ x > \frac{{ -
				a}}{b};b > 0}\\
	{{\boldsymbol{\beta }}{e^{{\bf{S}}x}}{{\bf{S}}^0}}&;&{{\rm{for }}\ x < \frac{{ - a}}{b};b
		< 0}
	\end{array}} \right.,
\]
 \footnote{Throughout the paper, if \textbf{A} is a matrix then \textbf{A}$^{0}$= -\textbf{Ae} being \textbf{e} a column vector of ones with appropriate
order} where \textbf{S}$^{0}$ is the column vector -\textbf{Se}=-b\textbf{Te}=b\textbf{T}$^{0}$.\\

 The moment-generating function is given by ${M_X}\left( t \right) =  - {\boldsymbol{\beta }}{\left( {{\bf{S}} + {\bf{I}}t}
	\right)^{ - 1}}{e^{ - \left( {{\bf{S}} + {\bf{I}}t} \right)a/b}}{{\bf{S}}^0},$ and
then $E\left[ {{X^n}} \right] = {\left. {\frac{{{\partial ^n}{M_X}\left( t
				\right)}}{{\partial {t^n}}}} \right|_{t = 0}}$.\\
			
From this expression the first and second moments are
\[
\begin{array}{l}
\ E\left[ X \right] =  - {\boldsymbol{\beta }}{e^{ - {\bf{S}}a/b}}{{\bf{S}}^{ -1}}{\bf{e}} - \frac{a}{b} \\
\ \\
\ E\left[ {{X^2}} \right] = \frac{1}{{{b^2}}}\left[ {2{\boldsymbol{\beta }}{{\bf{e}}^{ -{\bf{S}}a/b}}{{\bf{S}}^{ - 1}}\left( {\frac{1}{{{b^2}}}{{\bf{S}}^{ - 1}} + \frac{a}{b}{\bf{I}}} \right){\bf{e}} + {a^2}} \right]. \\
\end{array}
\]			
Consequently,
\[
Var\left( X \right) = \frac{1}{{{b^4}}}\left[ {2{\boldsymbol{\beta }}{e^{ -{\bf{S}}a/b}}{{\bf{S}}^{ - 2}}{\bf{e}} - {{\left( {{\boldsymbol{\beta }}{e^{ - {\bf{S}}a/b}}{{\bf{S}}^{ - 1}}{\bf{e}}} \right)}^2}} \right].
\]

 Let's see that the finite addition of independent PH distributions or homothecy of PH distributions is PH distributed.  \\

\noindent \textit{\textbf{Result 1 ({\color{Red}{Summation}} of independent {\color{Red}{PH}} distributions)}}\\
 Let $\left\{ {{Y_i};i = 1, \ldots ,n} \right\}$ be a finite sequence of independent PH distributions with representation $(\boldsymbol{\alpha}_i, \textbf{T}_i)$ for \textit{i}=1,...,\textit{n}. Then, the variable $W_n=\sum_{i=1}^nY_i$ is {\color{Red}{PH}} distributed with representation  $(\boldsymbol{\rho}_n,\textbf{L}_n)$ given by \\

${{\boldsymbol{\rho }}_n} = \left( {{{\boldsymbol{\alpha }}_1},{\bf{0}}} \right)$  and
${{\bf{L}}_n} = \left( {\begin{array}{*{20}{c}}
	{{{\bf{T}}_1}}&{{\bf{T}}_1^0 \otimes {{\boldsymbol{\alpha }}_2}}&{}&{}&{}&{}\\
	{}&{{{\bf{T}}_2}}&{{\bf{T}}_2^0 \otimes {{\boldsymbol{\alpha }}_3}}&{}&{}&{}\\
	{}&{}&{{{\bf{T}}_3}}&{{\bf{T}}_3^0 \otimes {{\boldsymbol{\alpha }}_4}}&{}&{}\\
	{}&{}&{}& \ddots & \ddots &{}\\
	{}&{}&{}&{}&{{{\bf{T}}_{n - 1}}}&{{\bf{T}}_{n - 1}^0 \otimes {{\boldsymbol{\alpha
			}}_n}}\\
	{}&{}&{}&{}&{}&{{{\bf{T}}_n}}
	\end{array}} \right),$ \\

\noindent {\color{Red}{where $\otimes$ is the Kronecker product defined as follows. Let $\mathbf{A}$ and $\mathbf{B}$ be two matrices with order $m \times n$ and $k \times l$ respectively. Then, $\mathbf{A} \otimes \mathbf{B}$ is a matrix with order $mk \times nl$ defined as $(a_{ij}\mathbf{B}).$}}

\noindent \textit{Proof.}\\
The proof of this result is developed through induction. It is well known that the distribution of $W_2$ is given by the convolution of $Y_1$ and $Y_2$. If we denote to the distribution function of $Y_i$ as $F_i$ then the distribution function of $W_2${\color{Red}{, convolution of $F_1$ and $F_2$, denoted by $*$,}} is

\[
{W_2}\left( t \right) = {F_1} * {F_2}\left( t \right) = \int_0^\infty
{{F_1}\left( {du} \right){F_2}\left( {t - u} \right)du} .
\]

 It is well-known that the Laplace-Stieltjes transform of the convolution is the product of the Laplace-Stieltjes transforms and that there is a biunivocal relationship between the original distribution and its Laplace-Transform.

Given the distribution function of a {\color{Red}{PH}} distribution with representation $(\boldsymbol{\alpha}_i, \textbf{T}_i)$, then its Laplace-Stieltjes transform is given by
\[
F_i^*\left( s \right) = {{\boldsymbol{\alpha }}_i}{\left( {s{\bf{I}} - {{\bf{T}}_i}}
	\right)^{ - 1}}{\bf{T}}_i^0.
\]
 Then,
\[
W_2^*\left( s \right) = {{\boldsymbol{\rho}}_2}{\left( {s{\bf{I}} - {{\bf{L}}_2}}
	\right)^{ - 1}}{\bf{L}}_2^0
= \left( {{{\boldsymbol{\alpha }}_1},{\bf{0}}} \right){\left( {\begin{array}{*{20}{c}}
		{s{\bf{I}} - {{\bf{T}}_1}}&{{\bf{T}}_1^0 \otimes {{\boldsymbol{\alpha }}_2}}\\
		{\bf{0}}&{s{\bf{I}} - {{\bf{T}}_2}}
		\end{array}} \right)^{ - 1}}\left( {\begin{array}{*{20}{c}}
	{\bf{0}}\\
	{{\bf{T}}_2^0}
	\end{array}} \right)
\]
\[
= \left( {{{\boldsymbol{\alpha }}_1},{\bf{0}}} \right)\left( {\begin{array}{*{20}{c}}
	{{{\left( {s{\bf{I}} - {{\bf{T}}_1}} \right)}^{ - 1}}}&{{{\left( {s{\bf{I}} -
					{{\bf{T}}_1}} \right)}^{ - 1}}{\bf{T}}_1^0{{\boldsymbol{\alpha }}_2}{{\left( {s{\bf{I}} -
					{{\bf{T}}_2}} \right)}^{ - 1}}}\\
	{\bf{0}}&{{{\left( {s{\bf{I}} - {{\bf{T}}_2}} \right)}^{ - 1}}}
	\end{array}} \right)\left( {\begin{array}{*{20}{c}}
	{\bf{0}}\\
	{{\bf{T}}_2^0}
	\end{array}} \right)
\]
\[
= {{\boldsymbol{\alpha }}_1}{\left( {s{\bf{I}} - {{\bf{T}}_1}} \right)^{ -
		1}}{\bf{T}}_1^0 \cdot {{\boldsymbol{\alpha }}_2}{\left( {s{\bf{I}} - {{\bf{T}}_2}}
	\right)^{ - 1}}{\bf{T}}_2^0 = F_1^*\left( s \right) \cdot F_2^*\left( s \right).
\]
 We assume that ${W_{n - 1}} = \sum\limits_{i = 1}^{n - 1} {{Y_i}} $ is
PH-distributed with representation $\left( {{{\boldsymbol{\rho }}_{n - 1}},{{\bf{L}}_{n -
			1}}} \right)$. Given that ${W_n} = {W_{n - 1}} + {Y_n}$ and  ${{\boldsymbol{\rho }}_n} =
\left( {{{\boldsymbol{\rho }}_{n - 1}},{\bf{0}}} \right)$ and  ${{\bf{L}}_n} = \left(
{\begin{array}{*{20}{c}}
	{{{\bf{L}}_{n - 1}}}&{{\bf{L}}_{n - 1}^0}\\
	{\bf{0}}&{{\bf{T}}_n^0}
	\end{array}} \right),$ then
\[
W_n^*\left( s \right) = {{\boldsymbol{\rho }}_n}{\left( {s{\bf{I}} - {{\bf{L}}_n}}
	\right)^{ - 1}}{\bf{L}}_n^0 = {{\boldsymbol{\rho }}_{n - 1}}{\left( {s{\bf{I}} -{{\bf{L}}_{n - 1}}} \right)^{ - 1}}{\bf{L}}_{n - 1}^0 \cdot {{\boldsymbol{\alpha
	}}_n}{\left( {s{\bf{I}} - {{\bf{T}}_n}} \right)^{ - 1}}{\bf{T}}_n^0 = W_{n -
	1}^*\left( s \right) \cdot F_n^*\left( s \right).
\] \\
\noindent \textit{\textbf{Corollary 1}}\\
Let $\left\{ {{X_i};i = 1, \ldots ,n} \right\}$ be a finite sequence of independent {\color{Red}{LPH}} distributions with PH-distributions associated given by
$\left\{ {{Y_i} = {a_i} + b{X_i};i = 1, \ldots ,n} \right\}$ with representation
$\left( {{{\boldsymbol{\alpha }}_i},{{\bf{T}}_i}} \right)$ for
\textit{i}=1,...,\textit{n}. Then, the variable $\Lambda_n=\sum_{i=1}^nX_i$  is {\color{Red}{LPH}} distributed with representation $\left( {\sum\limits_{i = 1}^n {{a_i}} ,b,{{\boldsymbol{\rho}}_n}{e^{{{\bf{L}}_n}\sum\limits_{i = 1}^n {{a_i}} }},b{{\bf{L}}_n}} \right).$\\

\noindent \textit{Proof.}\\
From result 1, $\sum\limits_{i = 1}^n {{Y_i}}  = b{\Lambda _n} + \sum\limits_{i = 1}^n {{a_i}}$ is PH with representation $\left( {{{\boldsymbol{\rho }}_n},{{\bf{L}}_n}} \right)$. Then, ${\Lambda _n} = \frac{1}{b}\sum\limits_{i = 1}^n {{Y_i}}  - \frac{1}{b}\sum\limits_{i = 1}^n {{a_i}} $ is {\color{Red}{LPH}} with representation $\left( {\sum\limits_{i = 1}^n {{a_i}} ,b,{{\boldsymbol{\rho}}_n}{e^{{{\bf{L}}_n}\sum\limits_{i = 1}^n {{a_i}} }},b{{\bf{L}}_n}} \right).$\\

 Next, we show that a positive homothecy of a PH distribution is also PH distributed. \\

\noindent \textit{\textbf{Result 2}}\\
 Let $Y$ be a {\color{Red}{PH}} distribution with representation $(\boldsymbol{\alpha}, \textbf{T})$ then the variable $\gamma Y$ is {\color{Red}{PH}} distributed with representation $(\boldsymbol{\alpha},\frac{1}{\gamma}\textbf{T})$ , being $\gamma$ a non-negative real number.

\noindent The proof of this result is immediate. Thus,
\[
 P(\gamma Y>t)= P(Y>t/\gamma)={\boldsymbol{\alpha }} e^ {\frac{1}{\gamma}{\bf{T}}t}  {\bf{e}} \ \ \ ; \ \ \ t > 0.
\] \\
\noindent \textit{\textbf{Corollary 2}}\\
 Let $X$ be a {\color{Red}{LPH}} distribution with representation $(a,b,\boldsymbol{\beta},\textbf{S})$, then the variable $\gamma X$ is {\color{Red}{LPH}} with representation $\left(\left|\gamma\right|a,b \cdot sgn(\gamma),\boldsymbol{\beta},\frac{1}{\gamma}\boldsymbol{S}\right)$, being $\gamma$ a non-zero real number, {\color{Red}{$|\cdot|$ the absolute value function}} and $sgn(\cdot)$ the sign function.\\

\noindent \textit{Proof.}\\
 If $X$ is a {\color{Red}{LPH}} distribution with representation  $(a,b,\boldsymbol{\beta},\textbf{S})$, then there exist $a$ and $b$ such that $Y = a + bX$ is PH$\left( {{\boldsymbol{\alpha }},{\bf{T}}} \right)$ where ${\boldsymbol{\beta}} = {\boldsymbol{\alpha }}{e^{{\bf{T}}a}}$ and ${\bf{S}} = b{\bf{T}}$.

 Then, from \textit{Result 2} we have that any homothecy of a LPH distribution is also LPH distributed.

\begin{itemize}
	\item If $\gamma  > 0$, $\gamma Y = \gamma a + b\gamma X$ is PH$\left( {{\boldsymbol{\alpha }},\frac{1}{\gamma }{\bf{T}}} \right)$. \\
	Then, \\
	$\gamma X$ is LPH with representation $\left( {\gamma a,b,{\boldsymbol{\alpha
		}}{e^{{\bf{T}}a}},\frac{b}{\gamma }{\bf{T}}} \right)\equiv \left( {\gamma a,b,{\boldsymbol{\beta}},\frac{1}{\gamma }{\bf{S}}} \right)$.
	\item If $\gamma  < 0$, $ - \gamma Y =  - \gamma a - b\left( {\gamma X} \right)$ is PH$\left( {{\boldsymbol{\alpha }},\frac{{ - 1}}{\gamma }{\bf{T}}} \right)$.\\
	Then, \\
	$\gamma X$ is LPH with representation $\left( { - \gamma a, - b,{\boldsymbol{\alpha}}{e^{{\bf{T}}a}},\frac{b}{\gamma }{\bf{T}}} \right)\equiv \left( {-\gamma a,-b,{\boldsymbol{\beta}},\frac{1}{\gamma }{\bf{S}}} \right).$
\end{itemize}
\noindent Therefore $\gamma X$ is LPH distributed with representation $\left(\left|\gamma\right|a,b \cdot sgn(\gamma),\boldsymbol{\beta},\frac{1}{\gamma}\boldsymbol{S}\right).$

\noindent \textit{\textbf{Result 3 (Density of the LPH class)}}\\
The set of LPH distributions is dense in the set of probability distributions defined on any half-line of real numbers. \\
\noindent \textit{Proof.}\\
This theorem is proved from the classical result for PH distributions: the set of PH distributions is dense in the set of probability distributions on the nonnegative half-line.
Let $W$ be a random variable defined on $w>c$ for any real number $c$. It is immediate that $W-c$ is defined on the nonnegative half-line. Then, there exists a variable $Y$, PH distributed, so closed as desirable to $W-c$. Therefore, the variable $X=Y+c$, which is LPH, approximates to the initial variable $W$.\\
A similar reasoning can be done for the case when $W$ is defined on $w<c$ for any real number $c$. In this case $-W+c$ is defined on $\mathbb{R}^+$. Then, there exists a variable $Y$, PH distributed, so closed as desirable to $-W+c$. For this case, the variable $X=-Y+c$, which is LPH, approximates the initial variable $W$.

\section{LPH modelling of functional PCA}

Let $X$ be a functional variable whose  observed values are curves, and let us assume that $X= \left\{ X\left( t\right) :t\in
T\right\}$ is a  second
order stochastic process, continuous in quadratic mean, whose sample functions
belong to the Hilbert space $L^{2}\left( T\right)$ of square
integrable functions with the usual inner product
$
\left\langle f,g\right\rangle=\int_{T}f\left(  t\right) g\left(
t\right)  dt,\;\forall f,g\in L^{2}\left(  T\right).
$

In order to reduce the infinite dimension of a
functional variable and to explain its
dependence structure by a reduced set of uncorrelated variables, multivariate PCA was extended to the functional case
\cite{Deville1974}. The functional principal
components (p.c.'s) are obtained as uncorrelated generalized linear combinations of the process variables with maximum variance (Var).
Then, the $j-th$ p.c. score is  given by
$
 \xi_{j}=\int_{T} \left(X \left(t\right)-\mu(t)\right)f_{j}\left(t\right)dt,
$
where the weight function or loading $f_j$ is the value of the argument $f(t)$ that maximizes de objective function with the corresponding constraints
$$
\left\{ \begin{array}{ll}
  Var \left [\int_T \left(X \left(t\right)-\mu(t)\right)f \left(t\right) dt \right ] & \\
  \mbox{subject to} \; \|f\|^2=1 \; \mbox{and} \;\int f_\ell\left(t\right) f\left(t\right) dt=0, \; \;\ell=1,\ldots,j-1. &
        \end{array} \right.
$$ 
It can be shown that the weight functions are the eigenfunctions of
the covariance operator $C.$ That is, the solutions to the
eigenequation $C(f_{j}) (t) = \int C \left(t,s\right)f_j \left(s\right)ds =
\lambda_{j} f_j (t),$
where $C \left(t,s\right)$ is the covariance function  and
 $\lambda_{j}=Var[\xi_{j}].$
Then, the process admits the following orthogonal representation (K-L expansion):
$$
X \left(t\right)=\mu (t) + \sum_{j=1}^\infty \xi_{j}f_j\left(t\right),
$$
with $\mu(t)$ being the mean function.
This principal component decomposition can be truncated providing
the best linear approximation of the sample curves in the least
squares sense $X^{q}\left(t\right)=\mu(t)+\sum_{j=1}^{q}\xi_{j}f_j\left(t\right), $
whose explained variance is given by $\sum_{j=1}^q \lambda_j.$

There are three main groups of rules for choosing the number of principal components. The first one consists of ad hoc rules-of-thumb that work very well in practice. The most used chooses a cut-off  of total variability, somewhere between  $90-95\%$, and  selects the smallest value of components for which this chosen percentage is exceeded. A graphical procedure, named scree-graph, consists of  ploting the number of components against the eigenvalues and retaining the number of components defining an ‘elbow’ in the graph. The second type of rules is based on formal tests of hypothesis and makes distributional assumptions, as multivariate normality, that are often unrealistic. The Bartlett’s test to decide if the last eigenvalues are equal can be sequentially used to find the number of components that are not noise. The third group consists of statistically based rules, most of which do not require distributional assumptions, based on computationally intensive methods such as cross-validation and bootstrapping. A detailed study on principal components selection rules can be seen in Chapter 6 in \cite{Joliffe2002}. 

The main objective of this work is to model the whole process from the random principal components. Given that PH distributions are dense in the non-negative probability distributions, we show that if the principal components are LPH distributed with the same scale parameter, then the one-dimensional distributions of the process are also LPH.\\

\noindent \textit{\textbf{Corollary 3}}\\
Let us assume that each principal component $\xi_j$ is LPH distributed with representation $(a_j,b \cdot sgn\left(f_j(t)\right),\boldsymbol{\beta}_j,\textbf{S}_j)$ for a real number $t$ and $j=1,\ldots,q$. Then,
the centered process $X(t) - \mu(t)$ is also LPH distributed  with representation $$\left( \sum\limits_{j = 1}^q {\left|{f_j}\left( t \right)\right|{a_j}},b,{\boldsymbol{\rho }}_{j} e^{{{\bf{L}}_{q}\sum\limits_{j = 1}^q {\left|{f_j}\left( t \right)\right|{a_j}} }},b {\bf{L}}_{q} \right),$$ with $\boldsymbol{\rho}_{q}=\left(\boldsymbol{\alpha}_1,\boldsymbol{0}\right)$ and
\[
{{\bf{L}}_{q}} = \left( {\begin{array}{*{20}{c}}
	{\frac{1}{{\left| {{f_1}\left( t \right)}
				\right|}}{{\bf{T}}_1}}&{\frac{1}{{\left| {{f_1}\left( t \right)}
				\right|}}{\bf{T}}_1^0 \otimes {{\boldsymbol{\alpha }}_2}}&{}&{}&{}&{}\\
	{}&{\frac{1}{{\left| {{f_2}\left( t \right)}
				\right|}}{{\bf{T}}_2}}&{\frac{1}{{\left| {{f_2}\left( t \right)}
				\right|}}{\bf{T}}_2^0 \otimes {{\boldsymbol{\alpha }}_3}}&{}&{}&{}\\
	{}&{}&{}& \ddots & \ddots &{}\\
	{}&{}&{}&{}&{\frac{1}{{\left| {{f_{{{\color{Red}{q}}} - 1}}\left( t \right)}
				\right|}}{{\bf{T}}_{{{\color{Red}{q}}} - 1}}}&{\frac{1}{{\left| {{f_{{{\color{Red}{q}}} - 1}}\left( t \right)}
				\right|}}{\bf{T}}_{{{\color{Red}{q}}} - 1}^0 \otimes {{\boldsymbol{\alpha }}_{{\color{Red}{q}}}}}\\
	{}&{}&{}&{}&{}&{\frac{1}{{\left| {{f_{{\color{Red}{q}}}}\left( t \right)} \right|}}{{\bf{T}}_{{\color{Red}{q}}}}}
	\end{array}} \right),
\]

\noindent where $|f_j(t)|$ is the absolute value of $f_j(t)$.
\noindent \textit{Proof.}\\
From Corollary 2, it is deduced that for a real number $t$, \\
if $f_j\left(t\right)>0$ then ${f_j}\left( t \right){\xi _j}$ is LPH with representation $ \left( {{f_j}\left( t \right){a_j},b,{{\boldsymbol{\beta}}_j},\frac{1}{{{f_j}\left( t\right)}}{{\bf{S}}_j}} \right),$\\
if $f_j\left(t\right)<0$ then $ {f_j}\left( t \right){\xi _j}$ is LPH with representation $ \left( {-{f_j}\left( t \right){a_j},b,{{\boldsymbol{\beta }}_j},\frac{-1}{{{f_j}\left( t\right)}}{{\bf{S}}_j}} \right).$

\noindent Then, from Corollary 1,  $\sum\limits_{j = 1}^q {{\xi _j}{f_j}\left( t \right)}$ is also LPH.\\

\section{Application}
The devices employed in this paper are composed of a metal-oxide-semiconductor stack whose metal electrode used was copper (200 nm thick), a dielectric 10 nm thick (Hf$O_2$) and a bottom electrode made of /Si-$n^+.$ The resistive memories were fabricated and measured at the Institute of Microelectronics of Barcelona (CNM-CSIC). The variability of these  devices is generated by an inherent  stochastic process that changes extremely the inner resistance of the device by means of  resistive switching physical mechanisms. The experimental data consist of a sample of current-voltage curves corresponding to the reset-set cycles associated with  the formation and rupture of a conductive filament that shorts the electrodes and changes drastically the device resistance.  From the mathematical viewpoint, the main objective here is to determine the current probability distribution at each voltage in  the reset process by means of the K-L expansion and the LPH distributions previously introduced .

 In this study, we have 232 reset curves denoted by $\lbrace I_i(v) \ : \ v\in[0,V_{i-reset}], i=1,…,232 \rbrace$ with  $ V_{i-reset}$ being the reset voltage.   Before applying FPCA to characterize the whole process through the K-L expansion, we must carry out some important previous steps proposed in \cite{Aguilera-Morillo19}. Briefly, this approach consists in synchronising all curves in the same interval due to the reset voltage is different for each curve, and using P-spline smoothing to reconstruct all reset curves since we only have discrete observations at a finite set of current values until the voltage reset for each curve. In this paper, the initial domain was transformed in the interval [0,1] and a cubic B-Spline basis of dimension 20 with 17 equally spaced knots and penalty parameter $\lambda=0.5$ was considered. Figure 1 shows all the smoothed registered curves in the interval [0,1], denoted by  $\lbrace I_i^*(u) \ : \ u\in[0,1], i=1,...,n \rbrace,$ and  the estimation of the mean function (red line).
\begin{figure}
	\centering
	\includegraphics[scale=0.5]{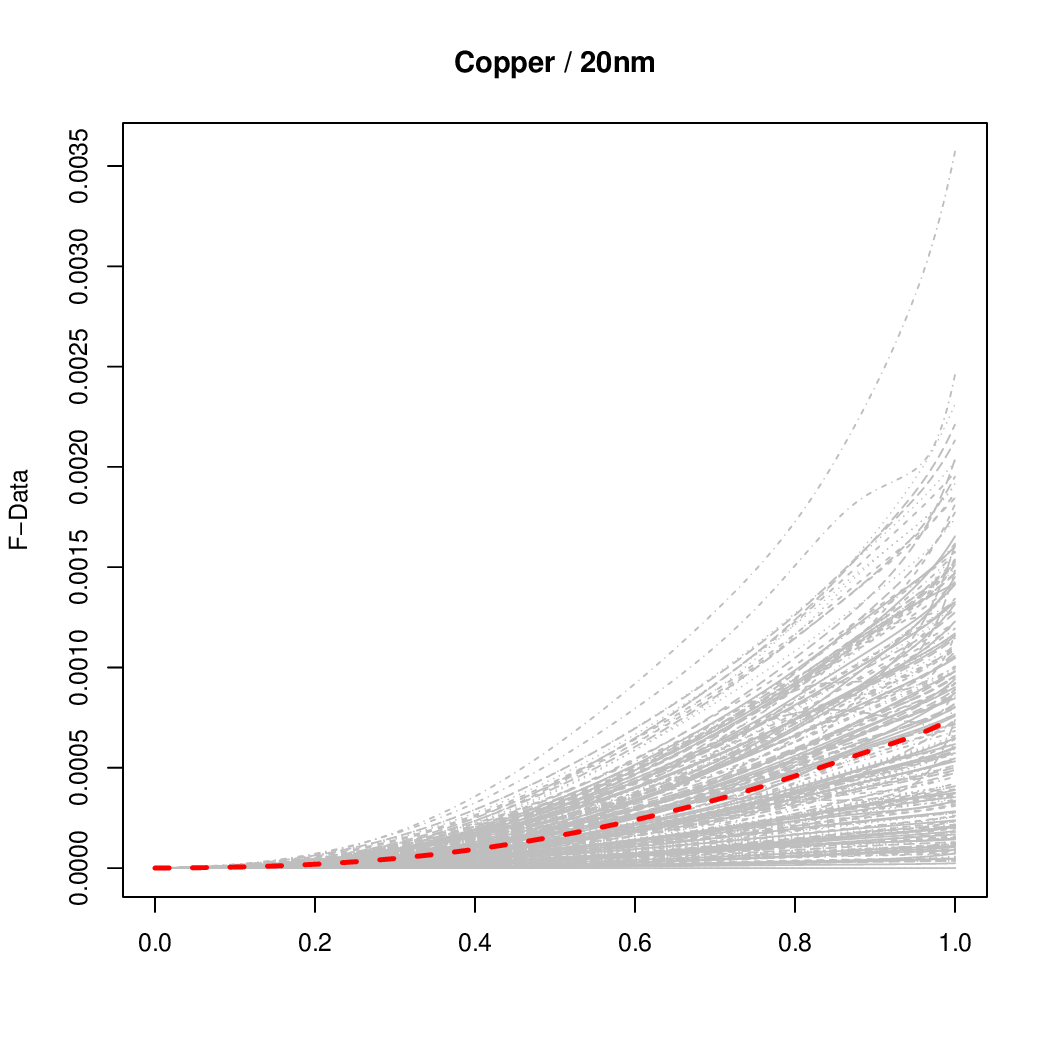}
	\caption{Sample group mean function (dashed red line) and all the P-spline smoothed registered curves. }
\label{Figure1}
\end{figure}

\begin{table}
	\begin{center}
		\begin{tabular}{c|c|c|c}
			\hline	
			Distribution & p-value K-S & p-value Anderson-Darling  & LogL \\
			\hline
			PHD & 0.11 & 0.054 & 18.11 \\
			Weibull & 0.004 & 0.004 & 0.78\\
			Normal & 0.02 & 0.006 & 7.79 \\
			Cauchy & $<0.001$ & $<0.001$ & -42.30\\
			\hline
		\end{tabular}
		\caption{Comparison among all distributions is considered. P-values of the Kolmogorov-Smirnov and Anderson-Darling tests, and the value of the maximum log-likelihood are showed for each distribution.}
		\label{Table1}
	\end{center}
\end{table}

Then, FPCA is estimated and the percentages of variance explained by the first four p.c.'s are 99.42, 0.44, 0.08 and 0.04, respectively. Let us observe that only the first p.c. explains more than 99\% of the total variability of the process. Hence, by considering the K-L expansion, principal component decomposition of the registered reset curves can be truncated in the first term as follows:
$
I{^*}^1(u)=\overline{I}^*(u)+\xi_1^*f_1^*(u), \ u \in [0,1].
$
This approach can be used for circuit simulation in this type of devices. Nevertheless, the probability distribution of the first p.c. is unknown.

In order to fit a probability model to the scores of the first p.c. different distributions were employed but none of them could be accepted (p-value associated with Kolmogorov-Smirnov test was $<0.01$ in all of them). Then,  some transformation is necessary. In this study, the LPH distributions associated with the linear transformation $1+1000\times \xi_1^*$ is considered. The p.c. is multiplied by 1000 because the standard variation of the principal component is very small, with minimum and maximum values of the component equal to $3.2e^{-04}$ and $7e^{-04}$, respectively. These facts produce a great number of phases in the corresponding PH distribution estimated (more than 1000 produced exploding effects). After that, all values were in $[-1,1]$ and then $1$ is added (to consider a PH distribution). Although the constant and slope values could be calculated by maximum likelihood (we are working on it), in this paper they were found {\it ad hoc,} taking into account that PHD are non-negative variables  (the values of the first p.c. are positives and negatives).

\begin{figure}
	\centering
	\includegraphics[scale=0.7]{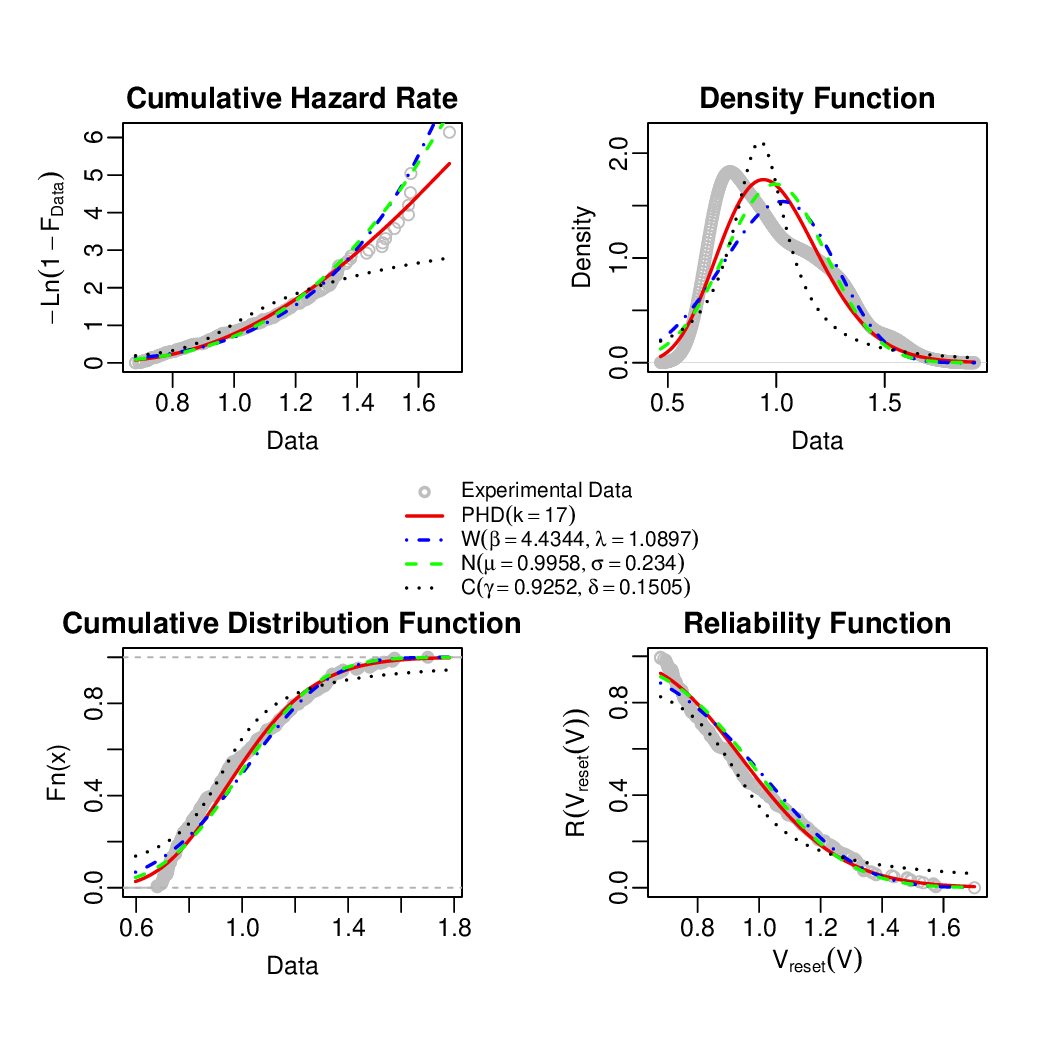}
	\caption{The cumulative hazard rate (topleft), the density function (topright), the cumulative distribution function (bottomleft) and the reliability function (bottomright) of experimental data with the fitting by means of PH, Weibull, Normal and Cauchy distributions.}
\label{Figure2}
\end{figure}

The EM algorithm  was used for estimating the parameters of a PHD with $m$ transient stages  and any internal structure for matrix \textbf{T} (\cite{Asmussen2000}\cite{Buchholz2014}). This methodology has also been applied to estimate the parameters of the PH distributions embedded in the study of the variability in resistive memories. The algorithm is described in \cite{Acal19}. The optimum value was reached for 21 stages. Besides, in order to prove that PHD is better than any other distribution, Weibull, Normal and Cauchy distributions were fitted as well. Their estimation by maximum likelihood are $W(\beta=4.4344, \lambda=1.0897)$, $N(\mu=0.9958,\sigma=0.234)$ and $C(\gamma=0.9252, \delta=0.1505),$ respectively. The results provided by all of them are given and compared in  Table \ref{Table1}. Thus, taking into account the logL value and the p-values of the K-S and the Anderson-Darling tests, the best distribution to get an accurate fit of the first p.c. score is the PH distribution. In fact, at 5\%  significance level, only the PH distribution can be accepted to model the first p.c. score according to the p-values provided by the Kolmogorov-Smirnov and Anderson-Darling tests. This conclusion can be achieved graphically. The cumulative hazard rate (topleft), the density function (topright), the cumulative distribution function (bottomleft) and the reliability function (bottomright) of data with the fitting by means of PH, Weibull, Normal and Cauchy distributions are displayed in Figure \ref{Figure2}.
In order to sum up, we have proved that the  considered linear transformation of the first p.c. is PH distributed with representation $(\boldsymbol{\alpha}, \textbf{T}).$  Therefore, the first p.c. score can be modelled through a LPH distribution with representation $(1,1000,\boldsymbol{\beta}, \textbf{S}).$ Finally, the reset process  $I{^*}^1(u)$ is LPH distributed as well with representation $$\left(\left|f_1^*(u)\right|-1000 \overline{I}^*(u) sgn\left(f_1^*(u)\right),1000  sgn\left(f_1^*(u)\right),\boldsymbol{\alpha}e^{\boldsymbol{T}\left(1-\frac{1000}{f_1^*(u)}\overline{I}^*(u)\right)},\frac{1000}{f_1^*(u)}\boldsymbol{T}\right).$$

\section{Conclusions}
A new probability distribution class with good properties, the LPH class, has been introduced to model the principal components in a matrix and algorithmic form. Multiple properties of this distribution class are developed, including that the LPH class is dense in the probability distribution class defined on any half-line of real numbers. Functional principal components analysis  provides a representation of a stochastic process through uncorrelated random variables called principal components. It is of great interest identifying the probability distribution of these components to analyse the random behaviour of the process. In this work, it has also been proved that the process, characterized through the K-L expansion, follows a LPH distribution at each point. The results have been applied to model the stochastic behaviour of resistive memories. In this case, one principal component is considered and the explicit representation of the LPH is given for the stochastic process at each point.

\section*{Acknowledgements}

We would like to thank F. Campabadal and M. B. Gonz\'{a}lez from the IMB-CNM
(CSIC) in Barcelona for fabricating and providing the experimental measurements of
the devices employed here. The authors thank the support of the Spanish Ministry of
Science, Innovation and Universities under projects TEC2017-84321-C4-3-R,
MTM2017-88708-P,  IJCI-2017-34038 (also supported by the FEDER program) and the PhD grant (FPU18/01779) awarded to  Christian Acal.
This work has made use of the Spanish ICTS Network MICRONANOFABS.


\bibliography{Biblio_engineering_revised}

\end{document}